\documentclass[prodmode,acmtap]{acmlarge}

\acmVolume{2}
\acmNumber{3}
\acmArticle{1}
\articleSeq{1}
\acmYear{2017}
\acmMonth{2}

\usepackage{graphicx}
\usepackage{listings}
\usepackage{amsmath}
\usepackage{stfloats,subcaption,hyperref}
\usepackage[flushleft]{threeparttable}
\usepackage[table]{xcolor}
\usepackage[ruled]{algorithm2e}
\definecolor{lightgray}{gray}{0.75}
\usepackage{fixme}
\fxusetheme{color}
\usepackage[titletoc,toc,title]{appendix}
\fxsetup{
     status=draft,
     author=,
     layout=inline, 
     theme=color
}
\definecolor{fxnote}{rgb}{0.8000,0.0000,0.0000}

\newcommand{\ie}{{\it i.e.}\hspace{0.1pc}}
\newcommand{\eg}{{\it e.g.}\hspace{0.1pc}}
\newcommand{\etal}{{\it et al.}\hspace{0.1pc}}

\SetAlFnt{\algofont}
\SetAlCapFnt{\algofont}
\SetAlCapNameFnt{\algofont}
\SetAlCapHSkip{0pt}
\IncMargin{-\parindent}

\title{DyAdHyTM: A Low Overhead Dynamically Adaptive Hybrid Transactional Memory on Big Data Graphs}
\author{Mohammad Qayum and Abdel-Hameed Badawy \affil{New Mexico State University}
Jeanine Cook
\affil{Sandia National Laboratories}
}

\begin{abstract}
Big data is a buzzword used to describe massive volumes of data that provides opportunities of exploring new insights through data analytics. However, big data is mostly structured but can be semi-structured or unstructured. It is normally so large that it is not only difficult but also slow to process using traditional computing systems. One of the solutions is to format the data as graph data structures and process them on shared memory architecture to use fast and novel policies such as transactional memory. In most graph applications in big data type problems such as bioinformatics, social networks, and cybersecurity, graphs are sparse in nature. Due to this sparsity, we have the opportunity to use Transactional Memory (TM) as the synchronization policy for critical sections to speedup applications. At low conflict probability TM performs better than most synchronization policies due to its inherent non-blocking characteristics. TM can be implemented in Software, Hardware or a combination of both. However, hardware TM implementations are fast but limited by scarce hardware resources while software implementations have high overheads which can degrade performance.  In this paper, we develop a low overhead, yet simple, dynamically adaptive (\ie at runtime) hybrid (\ie combines hardware and software) TM (DyAdHyTM) scheme that combines the best features of both Hardware TM (HTM) and Software TM (STM) while adapting to application's requirements. It performs better than coarse-grain lock by up to 8.12x, a low overhead STM by up to 2.68x, a couple of implementations of HTMs (by up to 2.59x), and other HyTMs (by up to 1.55x) for SSCA-2 graph  benchmark running on a multicore machine with a large shared memory.

\end{abstract}

\category{C.5.5}{Servers}{Clusters}[Symmetric Multiprocessors]
\terms{Dynamically Adaptive Hybrid Transactional Memory (DyAdHyTM)}
\keywords{Big Data, Graph Application, Synchronization, Hybrid Transactional Memory}

\begin{document}


\maketitle

\newpage
\section{Introduction}
\label{intro}


Though we are in the era of multicore and manycore processors era, we are not able to utilize the full computing power of all the cores due to challenges with critical sections in parallel workloads\cite{david2013everything}. To exploit all the cores in a system, programmers have to put significant effort to solve this issue of critical section with an appropriate synchronization first hand. A synchronization policy controls the access to shared data in a critical sections by multiple threads so that consistency and coherency of the data is maintained. The common synchronization policy used in shared data is coarse-grain synchronization (e.g. locks) that is easy to program, but not optimized as it unnecessarily sequentializes program execution due to its blocking characteristics. Examples of coarse grain synchronization based on locks include semaphores, mutexes, monitors, and barriers. These synchronization polices are used in conventional multi-threading programming libraries such as Pthread\cite{butenhof1997programming} or OpenMP\cite{dagum1998openmp}. The alternative is a fine-grain synchronization policy which is notoriously complex but provides better scalability due to its smaller granularity. For example, in CRAY processors, locks are used at the cacheline granularity using a full/empty bit\cite{bader2006designing}. This type of synchronization is not composable, which means locks cannot be combined in a modular fashion. In comparison to these kinds of conventional locking mechanisms, transactional memory provides easier programmability, better scalability, composabilty and sometimes lower overheads depending on workload behavior\cite{pankratius2011study}.

A collection of real-life big data problems that uses graph algorithms are already presented in the Graph500 benchmark~\cite{murphy2010introducing}.  Due to the sparse nature of data structures in most real word graph applications\cite{lumsdaine2007challenges}, TM will outperform most existing synchronization policies as it is inherently non-blocking\cite{kang2009efficient}. TM can be implemented in software, hardware, or in a combination of both that is hybrid. Hybrid Transactional Memory (HyTM) scheme optimizes the combination of Software Transactional Memory (STM) and Hardware Transactional Memory (HTM) depending on the requirement of the application\cite{damron2006hybrid}. A Dynamically Adaptive Hybrid Transactional Memory (DyAdHyTM) is a TM design that adapts to appropriate policies (\ie HTM or STM) at different phases of program execution on runtime based on the application's behavior. HTM is bound by limited hardware resources since micro-architecture implementation of TM policies can use up costly chip area\cite{dalessandro2011hybrid}. On the other hand, STM is limited by slow speed and high overheads due to high level abstractions implemented in Software\cite{matveev14reduced}. HyTM is even better approach than naive HTM or STM implementation since there is no best TM implementation for all applications. A HyTM not only combines fast HTM as the primary execution path and slow STM as the fall back execution path but also amends HTM with software extensions to co-operate with STM. Moreover, since synchronization is mostly executed in HTM due to adaptability of HyTM, few costly STM executions do not degrade performance much.

Currently, there are only a few commercial hardware implementations of transactional memory introduced by IBM and Intel. The most prominent implementation is the Transactional Synchronization Extensions (TSX)\cite{rajwar2012intel} implemented in the Intel 4th Generation Core\textsuperscript{TM} Processors. Restricted Transactional Memory (RTM)\cite{rajwar2012intel} in TSX is a new instruction set extension to the popular x86 instruction set architecture (ISA) comprising of four new instructions. For our DyAdHyTM, we used Intel's RTM  as best effort HTM due to its fastness and simple design. For STMs, we used GCC's STM due to its low overhead. Results with our designed DyAdHyTM show that it provides better performance than conventional synchronization methods such as locks, STMs and even HTMs for a standard graph application of large size as we scale the core/thread count, memory size and problem size. 

We tested our implementation on a large SMP machine of 28 cores and 64 GB memory for large scale graphs with up to 100s of millions of vertices and billions of edges to demonstrate that TM, and particularly, DyAdHyTM is a better synchronization scheme. The contributions of this paper can be summarized as follows:
\begin{enumerate}
   \item We develop a low overhead dynamically adaptive hybrid transactional memory that adapts to the workloads to optimize performance.
   \item We conduct a detailed performance comparison against various synchronization strategies including native HTM.
   \item We provide through statistics collected from the hardware insights into why the various techniques perform the way they preform.
\end{enumerate}

The rest of the paper is organized as follows: section~\ref{Sec:background} discusses TM in detail. Section~\ref{Sec:implement} describes our DyAdHyTM implementation and Section~\ref{Sec:results} discusses our experimental results. Section~\ref{Sec:related} discusses related works. Appendix~\ref{Sec:appendix} shows all the graphs in the paper in larger formatting.

\section{Background}
\label{Sec:background}

Transactional memory (TM) is a non-blocking synchronization scheme for shared regions that atomically updates memory (read or write to a chunk of memory). If it fails (i.e. memory should not have been updated), then TM rolls back to the previous state of memory. In TM, transactions (regions of code that use the TM paradigm) can work in parallel without being blocked. If any of the transactions conflict (i.e. several transactions try to access a shared data simultaneously) with other transactions it will abort and retry to execute later, but one of the transactions will commit (write updates to the memory) to guarantee progress.

In STM, all the transactional semantics and policies are implemented in software. However, it can also take advantage of hardware features such as atomic instructions. STM typically has high overheads but is flexible enough to suit different types of workloads, and it is also customizable. STM allows researchers to easily explore different TM designs. Furthermore, STM can be modified to accommodate changes in a compiler or OS, and it is relatively simple to integrate STM into existing system and language features. This is because STM has fewer intrinsic limitations than HTM, which is limited by cache size and cache hierarchy. 

Herlihy and Moss~\cite{herlihy1993transactional} proposed hardware transactional memory (HTM) in the early 90s. After two decades, Intel implemented the HTM (Intel's TSX protocol) in the Haswell processor family~\cite{hammarlund20134th}. In Intel's HTM, extensions to the microarchitecture and the instruction set are added to accommodate TM. The CPU handle the various TM policies \eg versioning, conflict detection, and conflict resolution. The hardware has to buffer old copies of the data in the caches or store buffers to allow for roll backs if need be. The coherence protocol is extended to accommodate such changes. The memory controller detects conflicts among transactions. New status registers are added to the CPU for performance evaluation/monitoring. A fallback mechanism is needed in case a transaction fails repeatedly to avoid a live-lock situation.

Typically, the overhead of the extensions and modifications to implement HTM is lower than that of an STM. Even with low overheads, HTM has some inherent limitations basically the bounded capacity of the CPU. In an HTM, the read and write sets must fit in the cache(s). This gives rise to a cache capacity limit. Also, cache associativity, hyperthreading, TLB size and replacement policies can limit the number of in-flight transactions in an HTM. Moreover, duration of transactions can be limited by hardware events \eg context switching, interrupts, exception handling, page-faults, and thread migrations. One of the solutions is to use HyTM to fall back to STM when transactions are aborted frequently due to HTM's limitations.

\subsection{Hybrid TMs}
A Hybrid Transactional Memory (HyTM) scheme uses a combination of STM and HTM depending on the application requirements~\cite{damron2006hybrid}. In most HyTMs, an HTM provides a fast execution path, while an STM serves as a fallback to handle situations where the HTM cannot execute the transaction successfully. Supporting the semantics of both HTM and STM is not trivial. For example, strong isolation can be lost if an HTM provides it whereas the STM does not. In HyTM, the hardware and software transactions not only coexist but also monitor common memory locations related to the status of a transaction. 

In a HyTM design, best-effort HTM~\cite{damron2006hybrid} is combined with the best applicable STM through supporting libraries. Best-effort HTM tries to execute as many transactions as possible within its implementation constraints (\eg cache size) but some transactions are bound to abort no matter what. Aborted and any remaining transactions are executed in via an STM that suits the application requirements.

Depending on how the HTM and STM are combined, there are three types of HyTMs that are common in the literature~\cite{matveev14reduced}, more details in Section~\ref{Sec:implement}:
\begin{enumerate}
  \item A fast path HTM with a slow STM as a fallback \eg Hybrid NOrec~\cite{dalessandro2011hybrid}, ASF~\cite{chung2010asf} and Reduced Hardware NOrec~\cite{matveev14reduced}
  \item HTM and STM in phases \eg PhTM~\cite{lev2007phtm}
  \item HTM and STM in parallel \eg Unbounded TM~\cite{ananian2005unbounded} and the proposed HyTM of Damron \etal~\cite{damron2006hybrid}.
\end{enumerate}

In a HyTM scheme, at any given time several HTM and STM transactions can work on the same shared region. However, this creates complexity in conflict detection and resolution due to the disparity between HTMs and STMs syntaxes. For example, issues arises such as how to handle a situation where an object (STM granularity) in an STM transaction conflicts with two cache lines (HTM granularity) of two HTM transactions but both HTM transactions are not conflicting with each other. Solving these issues and optimizing for different scenarios leads to highly sophisticated and complicated designs. Since graph applications are for the most part HTM bound due to small task size of critical sections~\cite{kang2009efficient}, a HyTM with simple, low overhead STM and best-effort HTM co-operation may suffice to provide good performance.

Most of the current HyTM designs take existing best-effort HTM from simulators or commercial implementations and add STM to them. Most of the complexity in HyTM is in implementing the software extensions. Therefore, there are many research opportunities for novel HyTM ideas such as TM extensions to higher level caches, a larger speculative store buffer, or using more hardware status registers to assist STM-HTM collaboration.

The sparsity of graphs in most real applications allows TM to outperform most existing synchronization techniques which are mostly blocking~\cite{CCWC}. In next section we discus why our HyTM approach is a better than any TM approach since it not only combines the best features of HTM and STM, but also adapts to the application requirements on runtime.

\section{Implementation} 
\label{Sec:implement}
\subsection{Adaptive HyTM}


An Adaptive Hybrid Transactional Memory (AdHyTM) is a TM design that adapts the different TM policies at different phases of program execution based on the behavior of the application. HTM is limited by the hardware resources depending on the particular microarchitecture implementation. On the other hand, STM is limited by slow speed due to high overheads since all the policies are implemented in high level abstractions in software. AdHyTM may be the best approach because there is no size fits all TM \ie no TM performs best for all applications. AdHyTM combines a fast HTM as the primary execution path and a slow STM as the fall back execution path. It also forces the HTM to cooperate with the STM to speed up synchronization of parallel applications. Since transactions are mostly executed in HTM, it is beneficial to make a few costly software extensions in the interface in order to efficiently fallback to a simple but slow STM path upon failure(s). For adaptability, AdHyTM can take several approaches on how HTM and STM can cooperate:
\begin{enumerate}
      \item Never run on HTM but run on STM if task sizes are large due to limited hardware resources
      \item Always run on HTM if task sizes are small
      \item Run cooperatively with HTM and STM if task sizes are not large and not small or over populates HTM. In that case, since HTM cannot execute all the transactions due to capacity limit, STM can execute the remaining transactions.
\end{enumerate} 

\subsection{Our Approach}
Our proposed HyTM also extends the HTM with a software extension in the interface to allow cooperation with the STM. The purpose of these extensions is to simplify the process of falling back to STM upon failure to execute in HTM.
For designing our AdHyTM, we considered following possibilities:
\begin{enumerate}
   \item One possible solution is to have a non-adaptable HyTM. In such a version, HTM will switch to STM when it fails to complete after a \textbf{fixed} number of retries.
   \item Another solution is to have a HyTM where HTM will switch to STM when it fails to complete after a \textbf{random} number of retries in stead of fixed retries.   
   \item A third solution is to tune the number of retries through manual (\ie static) profiling before falling back to STM for better performance~\cite{CCWC}.
	\item A fourth option is to design an STM friendly HTM. Since HTM provides flags describing transaction abort causes such as capacity limits, we can use these flags to adapt the HyTM to the application behavior.
\end{enumerate}

Researches have used the first three policies to design HyTMs~\cite{wang2012transactional,CCWC}. We applied a simplistic design based on lower overheads than complex algorithms to fallback to STM. The implementation in this paper uses the fourth step mentioned above that means it adapts dynamically on runtime to provide the optimum results. 

\begin{figure*}
\begin{subfigure}{0.49\linewidth}
\lstset{language=C}
\begin{lstlisting}[frame=single]
//tries set according to policy 
tries= RANDOM_RETRIES() or 
Fixed_NUM_RETRIES or Tuned_NUM_RETRIES
if HW__BEGIN begins successfully
  if (gbllock is locked)
    abort;
  else
    transactional code; 
    HW__COMMIT;
else if (tries >= 0)
  tries--; 
  retry in HW;
else //retrials quota ends
  atomic add(gblloc ,1); 
  SW__BEGIN
  if (no conflict in SW)
    transactional code; 
    SW_COMMIT;
  else 
    SW__ABORT; retry in SW;
    atomic sub(gblloc ,1);
return
\end{lstlisting}
\caption{RNDHyTM / FxHyTM / StAdHyTM\label{fig:StAdHyTM}}
\end{subfigure}%
\hspace{0.5 cm}
\begin{subfigure}{0.5\linewidth}
\lstset{language=C}
\begin{lstlisting}[frame=single]
tries= Fixed_NUM_RETRIES;
if HW__BEGIN begins successfully
  if (gbllock is locked)
    abort;
  else
    transactional code; 
    HW__COMMIT;
else if (tries >= 0)
  if (capacity limit reached)
    tries = 0; //switches to STM
  else
    tries--; retry in HW;  
else //retrials quota ends
  atomic add(gblloc ,1);
  SW__BEGIN
  if (no conflict in SW)
    transactional code; 
    SW__COMMIT;
  else 
    SW__ABORT; retry in SW;
    atomic sub(gblloc ,1);
return
\end{lstlisting}
\caption{DyAdHyTM\label{fig:DyAdHyTM}}
\end{subfigure}%
\vspace*{0.1in}
\caption{Random HyTM (RNDHyTM), Fixed HyTM (FxHyTM), Statically Adaptive HyTM (StAdHyTM) pseudo code (\subref{fig:StAdHyTM}) and Dynamically Adaptive  HyTM (DyAdHyTM) pseudo code(\subref{fig:DyAdHyTM})}
\vspace*{-0.06in}
\end{figure*}

\subsection{Random HyTM (RNDHyTM) Implementation}
In a RNDHyTM implementation as shown in the code listing in Figure~\ref{fig:StAdHyTM}, there is no dynamic adaptation. At first, a transaction tries to execute in HTM and if no STM transaction has captured the critical section, then it will commit. However, when a transaction fails, it tries assigned a number of possible retries in HTM. When its retrial quota ends, it will take a global lock and execute in STM.  The retrial quota is set with a random number ranges such as 1-20, 20-50, 50-100 etc. There is an overhead due to random number generation which is quite significant as we will discuss in section~\ref{Sec:results}. The whole algorithm is similar to DyAdHyTM which will be discussed in following section. 

\subsection{Fixed HyTM (FxHyTM) Implementation}
In FxHyTM implementation as shown in the code listing in Figure~\ref{fig:StAdHyTM}, there is no dynamic adaptation and the retrial number is fixed. It works in similar fashion to RNDHyTM. However, the retrial quota is set with a fixed random number such as 43, 23 or 76 without any design space exploration (DSE). Performance of this FxHyTM is unpredictable due to random retrial numbers.

\subsection{Statically Adaptive HyTM (StAdHyTM) Implementation}
In StAdHyTM implementation as shown in the code listing in Figure~\ref{fig:StAdHyTM}, there is an adaptation which is implemented statically. It works in similar fashion to RNDHyTM. However, the retrial quota is tuned. The retrial quota is tuned with design space explorations with different random number ranges, such as 1-20, 20-50, 50-100 etc. We explore the best range and then chose a fixed number from the range. The overhead is that we have to run the desired applications several times before hand to come up with tuned retrial numbers

\subsection{Dynamically Adaptive HyTM (DyAdHyTM) Implementation}
In our DyAdHyTM implementation, as shown in the code listing in Figure~\ref{fig:DyAdHyTM}, a transaction first tries to execute in HTM (\ie HW\_BEGIN). If no STM transaction has captured the critical section (\ie gblloc is free), HTM will commit (\ie HW\_COMMIT). If it fails to begin in HTM due to capacity limits, the number of retrials is forced to zero. In that case, transaction retries for the last time in HTM. If it fails, then it falls back voluntarily to STM without any retrials. For conflicts, or other reasons (\eg context switch), it retries in HTM for a fixed number of retrials. The number of retries (NUM\_RETRIES) is set to a fixed randomly (\ie similar to FxHyTM). When a transaction fails it gets assigned a number of possible retries. It takes a global lock and executes in STM (SW\_BEGIN). However, every time a transaction tries to enter the critical section in HTM, it will check for the global lock's availability (\ie if the critical section is already locked, the transaction will abort in HTM) which could be already acquired by another STM. The global lock can be captured by several STMs. When an STM takes the global lock, it increases its value by one. Therefore, when the value of the global lock decreases to zero, the HTM transactions can go forward. Since an STM transaction can conflict with other STM transactions, in that case, one of the STM transactions will succeed and commit (SW\_COMMIT) and the global lock's value will be restored (decreased). Even if an STM transaction fails, it restores the lock's value. Therefore, the lock will be released eventually when an STM transaction commits or aborts. There is no overhead of random number generator and running applications several times beforehand to have a tuned retrial number. Only overhead is to test the flags for abortion in runtime compare to FxHyTM.

\subsection{HTMs and STM Implementation}
\label{subsec:Implement:3HTMs}
We have also implemented three versions of HTMs. (1) HTM with atomic lock (HTMALock) similar to FxHyTM in  the code listing in Figure~\ref{fig:StAdHyTM}. Instead of an STM fallback, it fall backs to a lock based execution. However, when a transaction fall backs by using a lock, it waits for the lock to be free from other transactions before it can take the lock exclusively. This part is implemented with a while loop.  (2) The second version uses a spinlock and similar to previous HTM implementation. Hardware transactions frequently check the availability of the lock by spinning on the lock, while in the atomic lock version of HTM, hardware transactions atomically check for the availability of lock. (3) The last version is Intel's Hardware Lock Elision (HLE). The transaction first attempts in speculative mode and if it fails, in the next attempt it executes in non-speculative mode. This aborts other concurrent speculative transactions. We used the STM in GCC TM draft~\cite{schindewolf2009towards}.

\section{Experimental Results} 
\label{Sec:results}
In this section, we discuss the experimental  results of DyAdHyTM on the SSCA-2 benchmark~\cite{bader2006hpcs}. We test our DyAdHyTM against three versions of HyTM, three versions of HTM, an STM and coarse grain locking. We run the benchmark on an SMP machine that have 28 (with hyperthreading) cores backed by 64GBs of main memory. 

We choose the Scalable Synthetic Compact Applications graph analysis 2 (SSCA-2) benchmark~\cite{bader2006hpcs} since it is mostly used as a benchmark to test state of art TM implementations with the STAMP benchmark~\cite{minh2008stamp}. The benchmark has large integer operations, a large memory footprint, and irregular memory access pattern  that mimics real world large graphs. It has multiple kernels accessing a single data structure representing a weighted, directed multigraph. In addition to a kernel to construct the graph from the input tuple list, there are three additional computational kernels that operate on the graph. We only use two of the three kernels, namely, the generation kernel and the computation kernel. The generation kernel generates a large graph with scale as input. The generated graph is based on power-law~\cite{faloutsos1999power} and R-MAT format~\cite{chakrabarti2004r}. The computation kernel extracts edges by weight from the generated graph and forms a list of the selected edges.

This benchmark is readily available with an OpenMP version with its critical section having a built-in OpenMP lock. We modified the critical section to support STM, HTM(s) and HyTM(s). All experiments are conducted 20 times and the average execution time is reported. In comparing  DyAdHyTM with other state-of-the-art STMs, HTMs and HyTMs, most STMs and HyTMs~\cite{dalessandro2011hybrid,matveev2013reduced} have large overheads since they have more complexity in their designs which can take significant time and cost just to implement within the benchmark, and even then gains are not guaranteed. Therefore, we believe that low overhead STM and HTMs implementation on the native machine are the fastest schemes that we can compare against.

\begin{figure*}
\centering
  \begin{subfigure}{0.33\linewidth}
  \includegraphics[width=\linewidth]{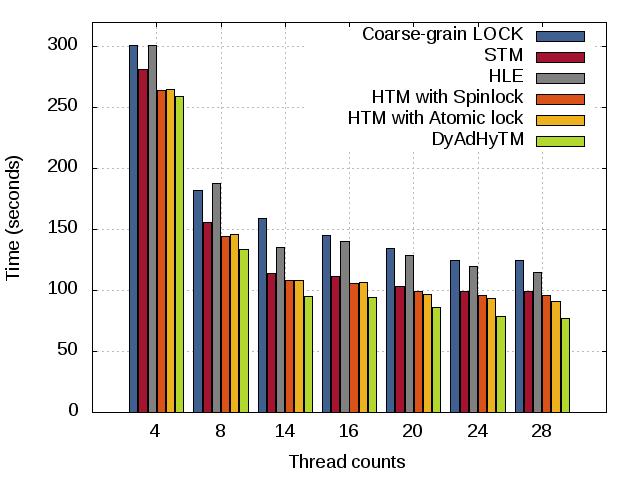}
  \caption{Scale 26: Two Kernels\label{fig:fig1}}
  \end{subfigure}%
  \begin{subfigure}{0.33\linewidth}
  \includegraphics[width=\linewidth]{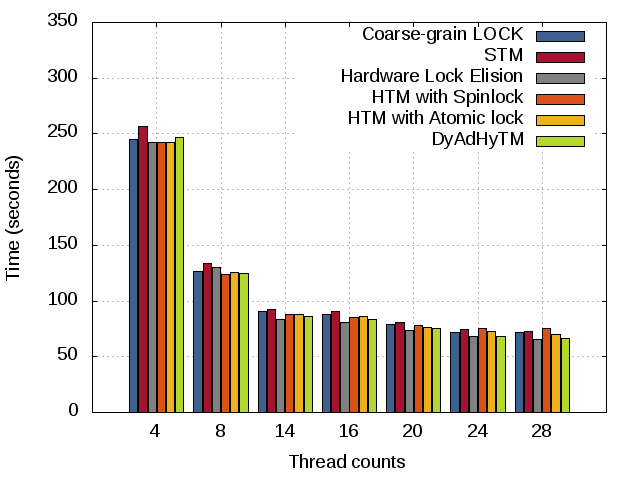}
  \caption{Scale 26: Generate \label{fig:fig2}}
  \end{subfigure}%
  \begin{subfigure}{0.33\linewidth}
  \includegraphics[width=\linewidth]{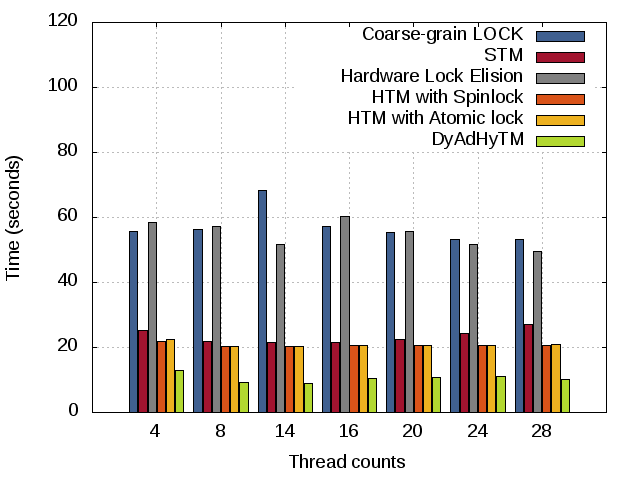}
  \caption{Scale 26: Compute\label{fig:fig3}}
  \end{subfigure}%
  \hfill
  \begin{subfigure}{0.33\linewidth}
  \includegraphics[width=\linewidth]{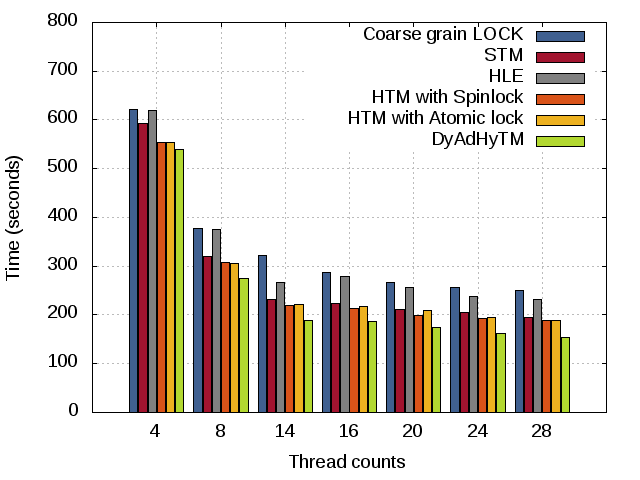}
  \caption{Scale 27: Two Kernel\label{fig:fig4}}
  \end{subfigure}%
  \begin{subfigure}{0.33\linewidth}
  \includegraphics[width=\linewidth]{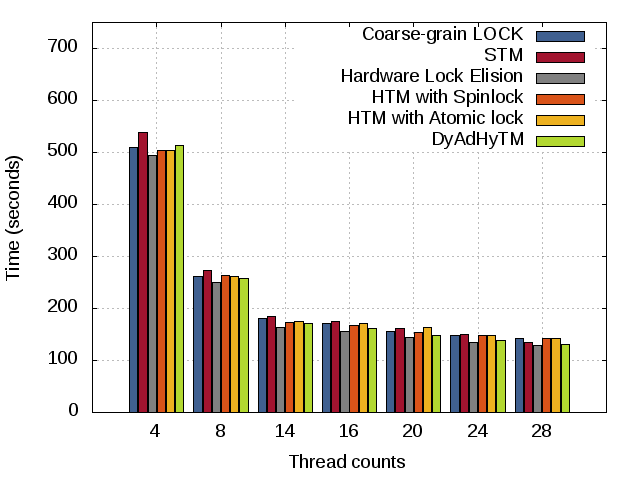}
  \caption{Scale 27: Generate\label{fig:fig5}}
  \end{subfigure}%
  \begin{subfigure}{0.33\linewidth}
  \includegraphics[width=\linewidth]{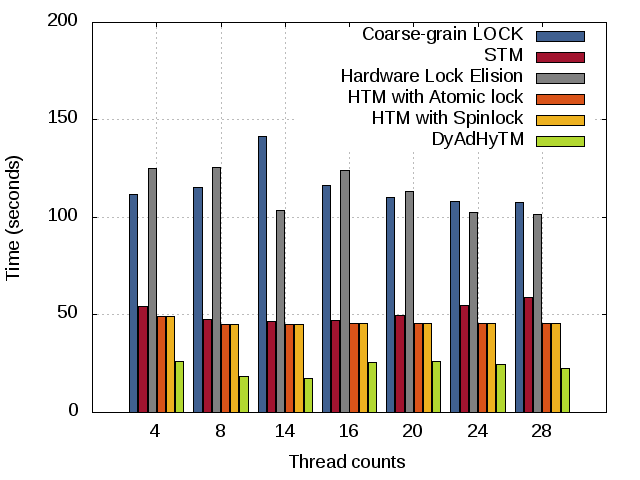}
  \caption{Scale 27: Compute\label{fig:fig6}}
  \end{subfigure}%
\caption{Performance improvement for 2 kernels (\subref{fig:fig1}) and (\subref{fig:fig4}), generation kernel (\subref{fig:fig2}) and (\subref{fig:fig5}), computation kernel (\subref{fig:fig3}) and (\subref{fig:fig6}) with DyAdHyTM  over coarse grain lock, STM, HLE, HTM for large scale 26 and 27 on a 28-core SMP node with 64 GBs memory, x-axis represents thread counts, y-axis represents execution time in seconds}
\label{fig:data3}
\end{figure*}

We conducted our experiments for larger scales (\ie 23 -- 27) with the SSCA-2 benchmark for all the policies on a workstation named ``Mickey". Mickey is a single SMP node with 14 cores or 28 logical cores if hyperthreading is enabled. It  has 64GBs of memory. The SMP node has a single Xeon Broadwell processor which has HTM implemented at the L1 and L2 caches. We show all our results for thread/core counts 4 -- 28 for better display purposes. It takes 2016.71 seconds with coarse grain lock for the two kernels with a single core but for 14 and 28 cores, it takes 321.50 and 250.52 seconds respectability.

Figures~\ref{fig:data3}(\subref{fig:fig1}) and~\ref{fig:data3}(\subref{fig:fig4}) show the total execution time of both graph generation and computation kernels of the SSCA-2 benchmark with all the synchronization policies for scales 26 and 27. Though, we have experimented with scales 23 -- 27, we only report results for 26 and 27 for space limitations. 

The results show that a simplistic STM implementation outperforms coarse grain lock for all scales and all thread counts. For scale 26, DyAdHyTM performs best among the policies with up to 1.62x speedup compared to coarse grain lock, 1.29x speedup compared to STM, 1.50x speedup compared to Hardware Lock Elision (HLE) version of HTM, and 1.18x speedup compared to the next best policy (\ie HTM with Spinlock at a thread count of 28). HLE's performance closely matches lock, we recall the reason from  section~\ref{Sec:implement} that HLE falls back to locked execution after first attempt of speculative execution.

Interestingly, for higher thread counts (\eg 20--28), HTMs performs similar to STM. We believe that some transactions are falling back to locks after they fail to execute in HTM because of conflicts. The overhead of aborts of HTM transactions, falling back overheads, the cost of serial execution in a lock are comparable to the overheads of STM since the only overhead in STMs is due to aborts due to conflicts of software transactions. 

The highest scale we can run on our target machine is 27. At this scale, most of the machine's memory (64GBs) is utilized for storing the graph (134 millions vertices and 1.03 billion edges). DyAdHyTM implementation gains 1.62x speedup compared to coarse grain lock and 1.29x speedup compared to STM in total execution time for both kernels at the largest thread count of 28 as is shown in Figure~\ref{fig:data3}(\subref{fig:fig4}). Both HTM policies outperform lock and STM. Finally, DyAdHyTM outperforms the next best policy (HTM with Spinlock) with 1.23x speedup. In DyAdHyTM, since both HTM and STM transactions can execute concurrently, we get better performance than just vanilla HTMs. However, very few HTM transactions fail to execute even after several retries and eventually, fall back to STM  although our goal is to maximize the number of transactions that succeed in committing in HTM. This is obviously the case since HTMs are the fastest. However, we can design an HTM without lock fallback that just keeps retrying with HTM, but its performance will be worse than falling back to a lock or STM. 

Figures~\ref{fig:data3}(\subref{fig:fig2}) and ~\ref{fig:data3}(\subref{fig:fig5}) show the results of a performance comparison of the  execution time of the graph generation kernel only of the SSCA-2 benchmark for all the policies. The graph generation kernel is a simple kernel with symmetric concurrency (\ie conflict probability is similar). Therefore, for all thread counts, most policies performs similarly. 

Figures~\ref{fig:data3}(\subref{fig:fig3}) and~\ref{fig:data3}(\subref{fig:fig6}) shows results for the computation kernel. This kernel posses dynamic conflict scenarios where threads compete against each others to update a critical section. For scale 27, (Figure~\ref{fig:data3}(\subref{fig:fig6})), the computation kernel at 14 threads, DyAdHyTM has a speedup of ~8.1x  compared to coarse-grain locking; and more than 2.5x compared HTM with spinlock. At thread count of 14, DyAdHyTM achieves the best execution time (\ie 17.442 seconds). Beyond 14 threads, we are using Hyperthreading (HT). In HT, the hardware threads share the caches, the micro-architecture and functional components, and therefore threads become more resource constrained and eventually become slower and transactions conflicts arise due to resource limitations. Therefore, beyond 14 threads, performance worsens. At 28 threads, overheads and conflicts kill the benefits of concurrency for the compute kernel. However, overall gain in total execution time for both kernels comes from the generation kernel since it takes 9x more time than the computation kernel (\ie the generation kernel dominates execution of the two kernels).  

\begin{figure*}
\centering
  \begin{subfigure}[b]{0.33\linewidth}
   \includegraphics[width=\linewidth]{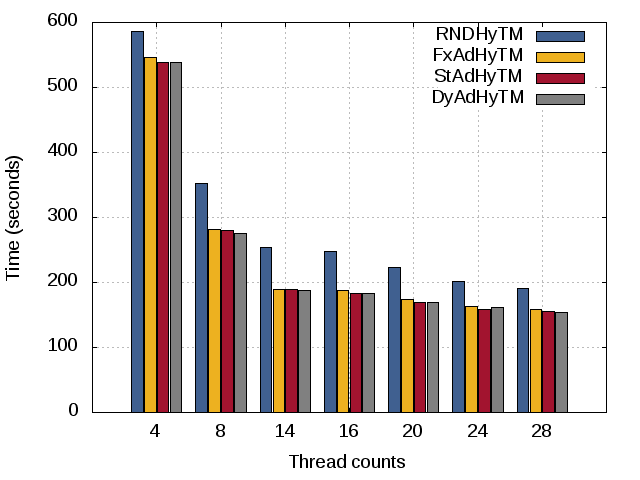}
  \caption{Scale 27: Two Kernels\label{fig:fig14}}
  \end{subfigure}%
  \begin{subfigure}[b]{0.33\linewidth}
  \includegraphics[width=\linewidth]{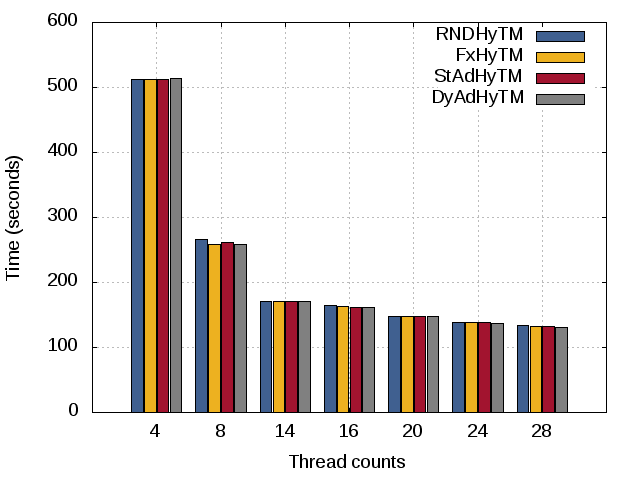}
  \caption{Scale 27: Computation\label{fig:fig24}}
  \end{subfigure}%
  \begin{subfigure}[b]{0.33\linewidth}
  \includegraphics[width=\linewidth]{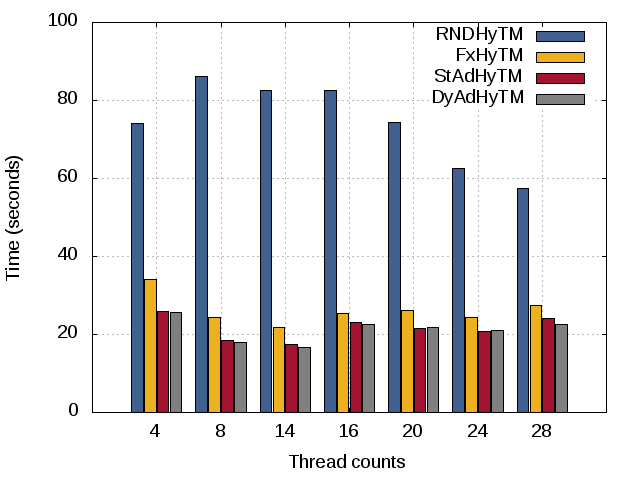}
  \caption{Scale 27: Computation\label{fig:fig25}}
  \end{subfigure}%
\caption{Execution time for two kernels (\subref{fig:fig14}), generation kernel(\subref{fig:fig24}), for computation kernel (\subref{fig:fig25}) for RNDHyTM, FxHyTM, StAdHyTM vs DyAdHyTM for large scale 27 on a 28-cores SMP node with 64 GBs memory  }
\label{fig:data3X}
\end{figure*}

Figures~\ref{fig:data3X}(\subref{fig:fig14}) and (\subref{fig:fig24}) show results for all HyTM policies for the two kernels, generation and computation kernels respectively. DyAdHyTM outperforms StAdHyTM by 1.4\%, FxHyTM by 3.81\%, and RNDHyTM by 24.8\% for two kernels for 28 threads. Recall that, FxHyTM has the lowest overhead but has unpredictable  performance and StAdHyTM has high unreported profiling overhead due to the manual tuning of the number of retries. For computation kernel and 28 threads, DyAdHyTM outperforms StAdHyTM by 4.2\%, FxHyTM by 21.8\% and RNDHyTM by 155.1\%. Due to the dominance of the generation kernel, the benefit are skewed. However, in a typical computation kernel on a graph, DyAdHyTM  gets the performance benefits since it is adaptive at runtime with low overhead.

\begin{figure*}
\centering
  \begin{subfigure}[b]{0.33\linewidth}
  \includegraphics[width=\linewidth]{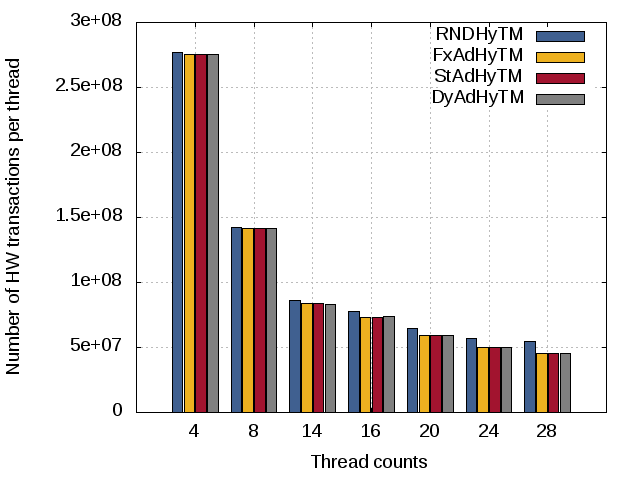}
  \caption{Scale 27: HTM transactions\label{fig:fig13}}
  \end{subfigure}%
  \begin{subfigure}[b]{0.33\linewidth}
  \includegraphics[width=\linewidth]{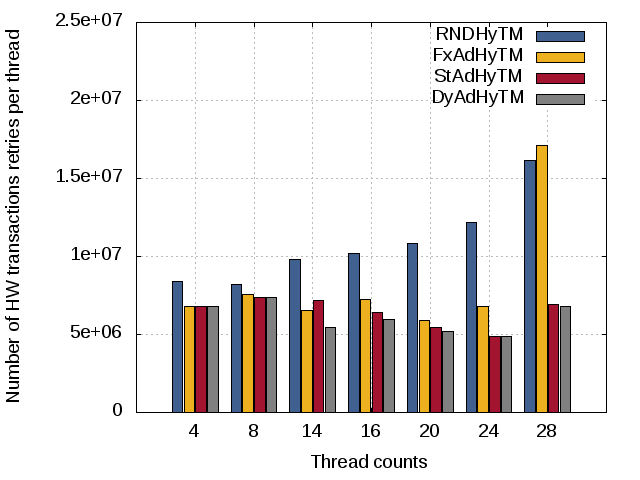}
  \caption{Scale 27: Retries\label{fig:fig23}}
  \end{subfigure}%
  \begin{subfigure}[b]{0.33\linewidth}
  \includegraphics[width=\linewidth]{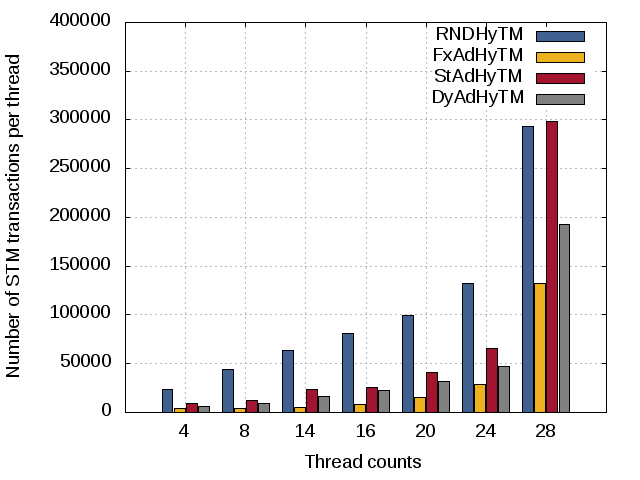}
  \caption{Scale 27: STM fallbacks\label{fig:fig33}}
  \end{subfigure}%
\caption{HTM Transactions per thread(\subref{fig:fig13}), HTM Retries numbers per thread(\subref{fig:fig23}), Number of STM transactions for RNDHyTM, FxHyTM, StAdHyTM vs DyAdHyTM for large scale 27 on a 28-cores SMP node with 64 GBs memory (\subref{fig:fig23}) }
\label{fig:data4X}
\end{figure*}

Figures~\ref{fig:data4X}(\subref{fig:fig13}) and (\subref{fig:fig23}) show the number of hardware transactions and retries per thread for RNDHyTM, FxHyTM, StAdHyTM and DyAdHyTM receptively. The number of retries for both StAdHyTM and DyAdHyTM is much smaller than RNDHyTM for all thread counts at scale 27. This explains that the speedup in both StAdHyTM and DyAdHyTM, come from higher commit success in HTM. In other words, HTM retries are lower as it is profiled for  StAdHyTM and adapts dynamically for DyAdHyTM. In case of FxHyTM, its number of retries is unpredictable since there is no intelligence in selecting the retries. For DyAdHyTM, it is further tuned by dynamically adapting to the executions by noticing causes of aborts and adopting retries accordingly. For example, at scale 27 and 28 threads, retries are 161.4M, 171M, 6.95M and 6.78M for RNDHyTM, FxHyTM, StAdHyTM and DyAdHyTM respectively. RNDHyTM takes a random number from 1--50 (Section~\ref{Sec:implement}). Also, DyAdHyTM has low overhead very close to FxHyTM since it only has to check status of HTM aborts.

Figure~\ref{fig:data4X}(\subref{fig:fig33})shows the number of STM transactions per thread for RNDHyTM, FxHyTM, StAdHyTM and DyAdHyTM. The number of STM executions for both FxHyTM, StAdHyTM and DyAdHyTM are much smaller than RNDHyTM for all the thread counts at scale 27. In case of 28 threads, DyAdHyTM wins but StAdHyTM's STM retries is close to RNDHyTM. Also, we are forcing HTM to switch to STM more frequently in StAdHyTM and DyAdHyTM, therefore,  STM counts are higher than FxHyTM. Gains mostly come from a less retries or less aborts in HTM.  Moreover, at higher thread counts, STM transactions per thread is much higher since conflicts arise as the number of threads increase but overall speedup comes from parallel execution of large number of threads.

\section{Related Work} 
\label{Sec:related}
The performance of most TM schemes depend on the implementation details. For example, most STM designs have special purposes behind their designs~\eg NOrec~\cite{dalessandro2010norec} which works on maximizing performance at low thread counts. 
SwissTM~\cite{dragojevic2009stretching} is optimized for high contention scenarios, and TinySTM~\cite{felber2010time} is a lightweight word-based STM implementation suitable for low contention scenarios. The TL2 STM~\cite{dice2006transactional} has high overheads but provides better scalability than NOrec. 

Most HTM implementations are agnostic of the applications' behaviors. Goel~\etal~\cite{goel2014performance} have experimented with HTM based on RTM similar to our HTM. They showed HTM scaling well for higher thread counts, with better execution times and lower energy consumption than TinySTM for eight threads. Also, our STM is a low overhead implementation of TinySTM. Christie~\etal~\cite{christie2010evaluation} have shown that AMD’s proposed Advanced Synchronization Facility (ASF) which is a simulator implemented HTM shows better scalability and performance than STMs such as  TinySTM for SSCA-2. However, ASF is not implemented in real hardware yet and comparing it to read hardware HTMs is not possible.

The very first papers that discussed HyTMs are Damron~\etal~\cite{damron2006hybrid} and Kumar~\etal~\cite{kumar2006hybrid}. Later on, other works included PhTM~\cite{lev2007phtm} which is based on the Sun ultrasparc HTM. Riegel~\etal~\cite{riegel2011optimizing} proposed a HyTM based AMD-ASF. Wang~\etal~\cite{wang2012evaluation} proposed another HyTM based on IBM Blue Gene/Q’s best-effort HTM with a Single-Global-Lock fallback. The most recent state of the art HyTMs are Hybrid NOrec~\cite{dalessandro2011hybrid}, Reduced Hardware-NOrec~\cite{matveev14reduced}, and Invyswell~\cite{calciu2014invyswell}. All of them are based on existing HTMs such as Intel's TSX or AMD's ASF with a single global lock. 

In-memory big data processing is a major area of research due to the availability of novel policies to utilize both large main memory and memory hierarchies as a data storage layer~\cite{zhang2015memory}. TM is one such policy that can be used efficiently for in-memory processing of big data applications. Kang and Bader~\cite{kang2009efficient} have applied STM for minimum spanning forest (MSF) problem. It is one of the earliest experiments with TM for graph  applications. Shang~\etal~\cite{shanggraph} have applied a two phase synchronization scheme in a graph application. They used coarse-grain lock (blocking) for high degree vertices and a non-blocking scheme for low degrees vertices. They assumed that in high conflict scenarios, non-blocking schemes may not work well. Our DyAdHyTM can be used in their non-blocking  implementations.

Our DyAdHyTM is based on Intel's TSX and adapts dynamically at run time. Its a low overhead solution that adapts to the available resources.  It is also designed with a single global lock for cooperation between HTM and STM. Most state of the art HyTMs are designed for general purpose applications and are tested with SSCA-2 benchmark mostly for small scales (less than 22) and with  low thread counts ($\leq  8$)~\cite{dalessandro2011hybrid,matveev14reduced,calciu2014invyswell}. Up to our knowledge, scalability for large graphs on HyTM was not considered before.

\section{Conclusion}
\label{Sec:conc}
There is no best synchronization policy that can fit for all purposes and applications. Researchers are looking for the appropriate synchronization policy for parallel applications that is scalable, provides easier semantic, and adaptive. Among the available policies, TM is one of the novel, well researched and rapidly evolving policies due to its properties of inherently non-blocking nature, availability in some commercial processors, ongoing research, and, best of all, ease of customization for a specific workload. Since most big data applications can be represented as large graphs which exhibits sparsity, TM is the logical solution for such concurrent applications. However, HTM limits task sizes due to it bounded resources while STM has speed limit due to high overheads. Therefore, we strongly believe that adaptive hybrid TM (HyTM) is a smart choice since it combines the best features of both HTM and STM, and can adapt to application requirements, specially to large scale graphs. Moreover, we designed a dynamically adaptive HyTM (DyAdHyTM) with little overhead which performs best among the all TM policies. Our DyAdHyTM provides speedup of up to 8.12x compared to coarse-grain lock for a large graph of 134M vertices and 1.03B edges on a manycore machine. 



\newpage
\bibliographystyle{spmpsci}
\bibliography{reference}              
\end{document}